\documentclass[12pt,preprint]{aastex}

\catcode`\@=11
\def\gapprox{\mathrel{\mathpalette\@versim>}}
\def\lapprox{\mathrel{\mathpalette\@versim<}}
\def\@versim#1#2{\lower2.45pt\vbox{\baselineskip0pt\lineskip0.9pt
    \ialign{$\m@th#1\hfil##\hfil$\crcr#2\crcr\sim\crcr}}}
\catcode`\@=12

\shorttitle{X-ray Point Sources in Galaxies}
\shortauthors{Colbert et al.}

\begin{document}

\title{Old and Young X-ray Point Source Populations in Nearby
Galaxies}

\author{Edward J. M. Colbert, Timothy M. Heckman, Andrew F. Ptak, 
and David K. Strickland}
\affil{Johns Hopkins University, Department of Physics and Astronomy,
Homewood Campus, 3400 North Charles Street, Baltimore, MD~~21218}

\and

\author{Kimberly A. Weaver}
\affil{Laboratory for High Energy Astrophysics, Code 662, NASA/GSFC,
  Greenbelt, MD~~20771; and Johns Hopkins University, Department of 
  Physics and Astronomy,
  Homewood Campus, 3400 North Charles Street, Baltimore, MD~~21218}

\begin{abstract}
We have analyzed Chandra ACIS observations of 
32 nearby spiral and elliptical galaxies and present the
results of 1441 X-ray point sources that were detected in these 
galaxies.
The total {\it point-source} X-ray (0.3$-$8.0 keV) luminosity L$_{XP}$ is well
correlated with the B-band, K-band, and FIR+UV luminosities of spiral host
galaxies, and is well correlated with the B-band and K-band luminosities
for elliptical galaxies.  This suggests an intimate connection between
L$_{XP}$ and both the old and young stellar populations, for which K
and FIR+UV luminosities are reasonable proxies for the galaxy mass $M$ and 
star-formation rate $SFR$.  We derive proportionality constants
$\alpha =$ 1.3 $\times$ 10$^{29}$ erg~s$^{-1}$~M$_\odot^{-1}$ and 
$\beta =$ 0.7 $\times$ 10$^{39}$ erg~s$^{-1}$~(M$_\odot$~yr$^{-1}$)$^{-1}$,
which can be used to estimate the old and young components from $M$ and 
$SFR$, respectively.  The cumulative X-ray luminosity functions for the
point sources have significantly different slopes.  For the spiral and
starburst galaxies, $\gamma \approx$0.6$-$0.8, and for the elliptical galaxies,
$\gamma \approx$1.4.  This implies that
{\it the most luminous point sources -- those with
L$_X \gapprox$ 10$^{38}$ erg~s$^{-1}$ --- 
dominate L$_{XP}$ for the spiral and
starburst galaxies.}
Most of the point sources have X-ray colors that are consistent with 
soft-spectrum (photon index $\Gamma \sim$ 1$-$2)
low-mass X-ray binaries, accretion-powered black-hole high-mass 
(BH HMXBs), or Ultra-Luminous X-ray sources (ULXs a.k.a. IXOs).
We rule out hard-spectrum neutron-star HMXBs (e.g. accretion-powered X-ray
pulsars)
as contributing much to L$_{XP}$.  
Thus, for spirals, L$_{XP}$ is dominated by ULXs and BH HMXBs.
We find no discernible difference between the X-ray colors of ULXs (L$_X \ge$
10$^{39}$ erg~s$^{-1}$) in spiral galaxies and point sources with 
L$_X \approx$ 10$^{38}$$-$10$^{39}$ erg~s$^{-1}$.
We estimate that $\gapprox$20\% of all ULXs found
in spirals originate from the older (pop~II) stellar populations, indicating
that many of the ULXs that have been found in spiral galaxies are in fact
pop~II ULXs, like those in elliptical galaxies.
We find that L$_{XP}$ depends {\it linearly} (within uncertainties) on both
$M$ and $SFR$, for our sample galaxies ($M \lapprox$ 10$^{11}$ M$_\odot$ and
$SFR \lapprox$ 10 M$_\odot$~yr$^{-1}$).
\end{abstract}


\section{Introduction}

As the resolution and sensitivity of X-ray imaging detectors continues to
improve, it becomes more and more feasible to study individual X-ray point
sources in external galaxies.  The point-spread function (PSF)
 of the Einstein IPC imaging spectrometer
($\sim$ 1$^\prime$) was typically too poor to distinguish emission from 
individual point sources in galaxies with distances $\gapprox$ 2 Mpc, and,
for more nearby galaxies, it was usually only sensitive to very luminous
sources with L$_X \gapprox$ 10$^{38}$ erg~s$^{-1}$.  The
ROSAT PSPC (PSF $\sim$ 20$^{\prime\prime}$) offered some improvement, and some
nearby galaxies could be studied in the soft (0.2$-$2.4 keV)
band.  Fabbiano (1989) gives a comprehensive review of X-ray sources 
in nearby galaxies in the pre-Chandra era.
The Chandra ACIS instrument has a tremendous improvement in both spatial
resolution (PSF $\sim$ 1$^{\prime\prime}$) and sensitivity, and covers the
bandpass 0.2$-$8.0 keV.
We are now able to study the properties of individual point sources in 
galaxies out to $\sim$20 Mpc, for which L$_X \approx$ 10$^{38}$ erg~s$^{-1}$
sources are detected in a reasonable (50~ks) exposure time.
Although the bright X-ray sources in the Milky Way (MW) are easily studied, many
of the MW point sources are viewed through the obscuring galaxy disk, so
one must correct for completeness when doing statistical work.  
Furthermore, in order to study how properties of classes of point sources 
correlate with host galaxy properties, it is imperative to study
large samples of different types of
galaxies.  We are now able to
study properties of low-mass X-ray binaries
(LMXBs), high-mass X-ray binaries (HMXBs), and Ultra-Luminous X-ray sources
(L$_X \ge$ 10$^{39}$ erg~s$^{-1}$; ULXs, a.k.a. 
Intermediate-luminosity X-ray Objects [IXOs]) 
in large samples of external galaxies with Chandra.  

Hard (kT $\sim$ 2$-$10 keV) X-ray emission is often present in the halos of
elliptical galaxies, and sometimes also in starburst galaxy nuclei
(e.g. Moran \& Lehnert 1997 and Ptak et al. 1997).
ASCA was not able to resolve or diagnose the origin of the hard emission
because of its poor spatial resolution (PSF $\sim$ 
1$^\prime$).
However, Chandra has
shown that X-ray {\it point sources} in spiral and elliptical galaxies often
emit a large fraction of the total hard X-ray luminosity.  We can now 
address some important questions with large surveys of galaxies with Chandra.
How do the X-ray point source populations of all of the different types
of X-ray binaries (XRBs)
and ULXs depend on the properties of the host galaxy?  What
is the exact connection between starbursts (or young star formation) and
XRB creation and evolution?
How does the host galaxy environment affect the evolution of the sources?
Is there a connection between the XRB population and the ULX population?  
Results from the many Chandra observations of spiral and elliptical galaxies 
already show important progress in many of these areas of study (e.g., see
review articles by Irwin et al. 2002a, Roberts et al. 2001, 
and Miller \& Colbert 2003, and references therein).

Here we present results from a study of the properties of the X-ray point 
source population in a sample of 5 Merger and Irregular galaxies
and 18 spiral galaxies, many of which have
high star formation rates (SFRs), and 9 elliptical 
galaxies, which typically have negligible SFRs.
We have analyzed the ACIS data for these 32 galaxies in exactly the same
way.
Thus, results from our processed data will not be affected by different
analysis techniques, as might have been the case if results from 
already published work on individual objects were used for the present 
study, such as tabulations listed in the many papers on Chandra observations
of galaxies (e.g, Eracleous et al. 2002, Swartz et al. 2003, 
Sarazin, Irwin \& Bregman 2001, and Irwin, Sarazin \& Bregman 2002b).
Since our calibration data, analyses software, or filtering 
criteria  may be different (see section 3.1), our
results (e.g. number of sources, count rates, fluxes, and luminosities)
may differ slightly from those already reported in the literature.

In section 2, we describe the galaxy sample.  Data reduction techniques are
discussed in section 3 and results are given in section 4.  We discuss the
implications of the results in section 5, and summarize in section 6.

\section{Description of Data and Galaxy Sample}

Data for twenty-seven of the galaxies in the sample were selected randomly
from the nearby (D $<$ 30 Mpc) NGC and Messier galaxies that were were
in the public Chandra archives on September 3, 2001.  Proprietary data
for galaxies NGC~3079, NGC~3628, NGC~5253 and NGC~4449 were also added.
When the Chandra observation of the very nearby elliptical galaxy NGC~3379
became public in February 2002, we added it as well.

Since most of the data were taken from the Chandra archives and
therefore the sample is already biased toward galaxies that are 
historically interesting in the X-ray band, there is no obvious
way to select a `complete,' unbiased sample.  Therefore, our goal was to 
select $\sim$30 galaxies
of all morphological types, and with a wide range in ``starburst
power,'' since we are interested in studying how the X-ray point source
population varies with host galaxy properties.
Since the XASSIST software (see section 3.1) typically requires $\sim$1
day of computer processing (or re-processing) per dataset, 
a sample size of $\sim$30 galaxies was both feasible, and large enough for
useful statistical work.  
We imposed the restriction that all of the sample galaxies must have 
X-ray luminosity sensitivity t/D$^2 >$ 0.044 ks~Mpc$^{-2}$, where $t$ is
the exposure time after filtering (see Table 1), and $D$ is the distance
to the galaxy (Table 2).
This corresponds to a luminosity sensitivity of $\approx$10$^{38}$
erg~s$^{-1}$ for a back-illuminated (BI) CCD, such as CCD~7.

Since our intent is to study the relationship between the X-ray point source
populations and the properties of the stellar component of the host galaxies
(and not the active super-massive black holes [SMBHs] and their possible
connection with intermediate-mass black holes [IMBHs], for example),
we intentionally omitted
the three powerful active
galaxies NGC~1068, NGC~4151, and M87.  Although
the peculiar elliptical galaxy Cen~A has an AGN with a radio jet, like M87,
it is the only ``elliptical'' galaxy with significant star formation, so we
decided to keep it in our sample.
LINERs and weak Seyferts were not omitted intentionally (see column 4 of
Table 2).  The composite starburst/AGN galaxy NGC~4945 was also kept.

Our final sample of 32 galaxies (see Table 1)
includes 28 galaxies from the archive, and proprietary data for four galaxies
(NGC~3079, NGC~3628, NGC~4449, and NGC~5253).

In Table 2, we list some host galaxy properties that are related to
the stellar 
content of the galaxy, in particular properties that are related to
the star-formation rate (SFR) and the
stellar mass (M).  The far-infrared (FIR) and far-ultraviolet luminosities
are proxies for the SFR, while the B-band and K$_s$-band luminosities are
proxies for the mass (e.g. Kennicutt 1998, Bell \& de~Jong 2001).
Note that the galaxies in our sample span a wide range in both morphological type, 
and in SFR/M (see L$_{FIR}$/L$_{B}$ or L$_{FIR+UV}$/L$_{K}$; Table 2).

For galaxies with recessional velocities $<$ 1000 km~s$^{-1}$, as listed in
the Third Reference Catalog of Bright Galaxies (RC3; de~Vaucouleurs et al. 1991),
distances were taken directly from Tully (1988).  We used 21-cm recessional
velocities when available, otherwise optical recessional velocities. 
For galaxies with recessional velocities $\ge$ 1000 km~s$^{-1}$, we 
calculated the distance using H$_0 =$ 75 km~s$^{-1}$~Mpc$^{-1}$.  The
distances to NGC~4038 and NGC~5094 were used for X-ray sources in 
NGC~4038/9 and NGC~5194/5, respectively.

\section{Data Reduction}

\subsection{X-ray Data}

All of the public ACIS data were retrieved from the SAO Chandra archives.
All of the data 
were reprocessed and reduced using a modified version of the XASSIST 
v0.757 (Ptak \& Griffiths 2003) scripts,
LHEASOFT v5.2, and CIAO v2.2.

Level-1 event files were reprocessed to level-2 event files with CIAO, using XASSIST.
XASSIST uses the basic data reduction steps recommended by the CXC ``threads.''
The optional 0.5 pixel position randomization was not performed.  After this reprocessing
step, each CCD is treated as a different detector, and individual CCD data
are processed separately.  
For data processing, PI values were constrained to be in the range 14$-$548, 
corresponding to
the energy range 0.2$-$8.0 keV, although fluxes and counts for X-ray colors
were computed from the data in the 0.3$-$8.0 keV range.

The CIAO {\sc wavdetect} source detection routine was then used on the reprocessed
level-2 event data to produce a preliminary list of point sources.
For all sources with more than 10 net counts, we further tested the robustness of the 
detection using a CPU-intensive 2D Gaussian image-fitting algorithm.
If a reasonable fit to the image was obtained with a Gaussian (source)
plus sloping plane (background) model, we were able to estimate a more
accurate count rate, and better estimates of the major and minor axes
of the source.  If the ratio of the major to minor axis was larger than 2.0,
the source was initially flagged as `extended.'
Likewise, if the size of the major or minor axes was larger than 
the PSF (at that off-axis angle), the source was also flagged as `extended.'
Sources detected at S/N $<$ 2.0 were rejected, as were sources with fitted
Gaussian sizes too small to be consistent with the ACIS PSF.

We then filtered only those sources inside the galaxy major and minor R$_{25}$ 
ellipse, as listed in RC3.  Each source was then individually inspected on-screen, 
and re-classified as `point-like' or `extended.'  
Any X-ray sources that were within
5$^{\prime\prime}$ of the galaxy nucleus (as listed in NED) were considered potential
active nuclei and were flagged for further examination.  After checking each case by
hand, a single nuclear X-ray source was omitted in each of the following seven galaxies:
NGC~3079, NGC~4258, NGC~4374, NGC~4579, NGC~4945, NGC~5128, and NGC~5194.
The `jet' sources in NGC~4258 and Cen~A were also
rejected.  We also rejected all of the artificial X-ray sources that were produced along
the readout column of high count-rate sources (NGC~4579 and 
NGC~5128).
After screening, we found a total of 1441 point sources in 32 galaxies 
(see Table 1 and Appendix).
As mentioned in section 1, our data reduction techniques and filtering methods
may be different from those of other workers, so our point source lists
may also differ.

The major and minor axes for elliptical source regions were determined 
directly from the Gaussian fitting, or from
{\sc wavdetect} if the Gaussian fitting was not performed, or failed.
Local annular regions surrounding the source, and centered on the source 
position, were used for background.
In crowded regions, when other source regions overlapped with the background
region, the contaminating part(s) of the background region were omitted.

These source and background regions were used directly 
to compute counts in three energy ranges, for hardness ratios
(see Table A1).  
X-ray fluxes in the 
0.3$-$8.0 keV band were determined for each source using the net count rates 
from the Gaussian fitting algorithm when available.  Otherwise, the count
rates from {\sc wavdetect} were used.
We converted the count rates to fluxes and luminosities 
using a simple power-law model with 
$\Gamma =$ 1.8, and the Galactic Hydrogen columns and distances listed in 
Table 2.
Auxiliary Response Files (ARFs)
generated using CIAO {\sc psextract} were used to compute fluxes, when
they were available.  Otherwise, on-axis ARF files were used.

We can estimate the uncertainties in our X-ray point-source luminosities using
a simple absorbed power-law model in XSPEC.
For a typical Galactic column of N$_H =$ 3 $\times$ 10$^{20}$ cm$^{-2}$, the
flux/count-rate ratio varies by 5$-$10\% when $\Gamma$ varies by $\pm$0.2 
($\Gamma =$ 1.6$-$2.0).  If $\Gamma$ varies by $\pm$0.7 ($\Gamma =$ 1.1$-$2.5), the
ratio varies by 30$-$40\%.  Uncertainties in the absorption column are also important.
Compared with a model with N$_H =$ 3 $\times$ 10$^{20}$ cm$^{-2}$, a more higly absorbed
model with N$_H =$ 3 $\times$ 10$^{21}$ cm$^{-2}$ has flux/count-rate ratios 5$-$25\% higher.
This would cause our quoted luminosities to underestimate the true luminosities.  For
an highly-absorbed source with N$_H =$ 3 $\times$ 10$^{22}$ cm$^{-2}$, the correction
is severe ($\approx$100$-$190\%).  Thus, unless the source is highly absorbed, our estimates of the 
observed X-ray luminosities are within $\approx$30$-$40\% of their actual value.

The total galaxy point-source X-ray luminosity L$_{XP}$ was then computed by 
summing all of the individual X-ray luminosities for each
of the point sources (see Tables A1 and 3).
We list in Table 3 
the fraction of L$_{XP}$ at 
L$_X \ge$10$^{38}$ erg~s$^{-1}$, and the fraction at
L$_X \ge$10$^{39}$ erg~s$^{-1}$ 
(i.e., the ULXs).
We also list the fraction of the total {\it number} of sources, and the 
total number of sources in these two high-luminosity ranges.
The relatively small scatter in the ratio of L$_{XP}$ to L$_K$ in the last 
column of Table 3 shows the strong correlation between the stellar light
and the X-ray light.
As discussed in section 3.2, we believe our values of L$_{XP}$ approximate
the actual
total point source luminosity within $\sim$20\% for most of our galaxies.
The exceptions are M82, NGC~253, and the elliptical galaxies, for which we
estimate an uncertainty of 40\%.

Hardness ratios were computed 
for all sources (Table A1), but we used only the sources
with $\ge$ 20 net counts for our analyses.
A total of 1017 sources met this criterion.  Most (810) of these sources
were detected on BI CCD~7.  One additional source
was detected on BI CCD5, and the rest of the sources were detected on 
front-illuminated (FI) CCDs (0, 1, 2, 3 and 6).

We selected three energy bands for computing hardness ratios: S 
(soft, 0.3$-$1.0 keV), M (medium, 1.0$-$2.0 keV), and H (hard, 2.0$-$8.0 keV).
Counts were extracted directly from the source and background event files
described above.
Hardness ratios of the form HR $=$ (C2$-$C1)/(C2$+$C1) were computed, 
where C1 and C2 are the net counts in the lower and higher energy bands, respectively.  
Hardness ratios from sources detected on the FI CCDs were 
transformed to the corresponding ratio for BI CCDs using
effective area (EA) curves from the CXC 
website\footnote{%
URL http://asc.harvard.edu/cal/Acis/Cal\_prods/effarea/4\_99/,
EA $*$ QE $*$ filt.trans. curve files files orbit\_i.dat [FI] and orbit\_s.dat [BI]
}.
Corrections to the hardness ratio were
obtained by estimating the correction to the counts in each energy band.
For each of the sources on the FI CCDs, we 
first fit a power-law model to spectra in each energy band.
We then computed the correction factor $f$ to the FI counts using the best-fit photon
index for that band $\Gamma_{band}$ and the equation
$$f_{band} = {{\int_{band} E^{-\Gamma_{band}}EA(BI) dE}\over{\int_{band} E^{-\Gamma_{band}}EA(FI) dE}}.        \eqno(1)$$
The corrected counts are then C1$^\prime = f_1$C1 and C2$^\prime = f_2$C2,
where $f_1$ and $f_2$ are the correction factors for the lower and high
energy bands, respectively.
Corrected hardness ratios HR$^\prime$ were then computed from 
HR$^\prime$ = ($\phi +$ HR) / (1 $+ \phi$\ HR), where $\phi =$ (1 $- f_1/f_2$)/(1 $+ f_1/f_2$).
Corrected hardness ratios for the sources are listed in Table A1.

\subsection{Unresolved X-ray Point Sources and Diffuse Hard X-ray Emission}

In order to test the accuracy of L$_{XP}$ due to the omission of blended
point sources, the erroneous addition of spurious point sources, and
``compact'' clumps of diffuse emission, we computed the total hard
(2.0$-$8.0 keV) counts from the individual point sources and compared
with the total hard counts from the images
within the R$_{25}$ ellipse.  Hard counts
are expected only for point sources (e.g., XRBs) and diffuse hard X-ray
emission, since the AGN and jet sources were omitted.  In general,
the hard counts for each method were consistent within $\sim$20\%,
which is typical of the uncertainty in the total hard counts due to the
large number of background counts in the ellipse.  For M82 and NGC~253,
there were excess hard counts, either in diffuse hard emission,
or blended ``unresolved'' point sources (both are edge-on spiral galaxies).
We also found evidence for excess hard counts in the elliptical galaxies.  This
is expected, since they are well known to have diffuse hard X-ray emission.
The diffuse hard emission in elliptical
galaxies could be from multiple unresolved XRBs, or could be emission
from diffuse, hot gas (e.g., Irwin et al. 2002b and references therein).

\subsection{Host Galaxy Luminosities}

As mentioned above, we calculated total luminosities from the galaxies in 
four separate energy bands: FIR, NIR K$_s$-band, optical B-band, and the 
far-ultraviolet (FUV).  All fluxes were converted to luminosities using the
distances listed in Table 2.  Far-infrared fluxes over the 40$-$120 $\mu$m
band were derived from the 60$\mu$m and 100$\mu$m IRAS fluxes using the 
method of Fullmer \& Lonsdale (1989).   
Near-infrared K$_s$-band spectral fluxes F$_\lambda$ 
were calculated from 20 mag~arcsec$^{-2}$ isophotal magnitudes listed in the 
2MASS Large Galaxy Atlas (Jarrett et al. 2003).  Optical B-band spectral fluxes
F$_\lambda$ were calculated from B$_T^0$ magnitudes (or m$_B^0$ magnitudes for
NGC~4038/9) listed in RC3.  For NGC~4038/9 and NGC~5194/5, the total spectral
flux was calculated using the sum of F$_\lambda$ for each galaxy.  In order
to quote all luminosities in roughly the same size band, we have calculated
L$_{K}$ and L$_B$ as $\lambda$L$_\lambda$.

Far-ultraviolet fluxes were estimated using SED-fitting to broad-band
total optical magnitudes (usually UBVRI, but in some cases UBVR or UBV),
from Prugniel \& Heraudeau (1998).  
We experimented with a wide range of different SED
databases, and obtained the best results using the observationally-based
SEDs of Kinney \& Calzetti (Kinney et al. 1996, Calzetti et al. 1994)
and the simulated HYPERZ SEDs (cf. Bolzonella et al. 2000).
There was generally good agreement between 
our estimated FUV fluxes and published measurements
(e.g., Rifatto et al. 1995, Marcum et al. 2001).
The best-fit SEDs were then
integrated from 1500$-$3500\AA\ to obtain the FUV flux.
We list the B, K, FIR and FUV luminosities for each galaxy in Table 2.

\section{Results}

\subsection{Simple Correlations of Total X-ray Point Source Luminosity with Host Galaxy Properties}

In Figure 1, we plot the total {\it point-source} X-ray luminosity
L$_{XP}$
against the optical B-band, NIR K-band, FIR and FIR$+$FUV luminosities of 
the host galaxy.
As noted by Fabbiano et al. (1988), both spiral and elliptical galaxies show a 
good correlation between the X-ray and B-band luminosity, suggesting that the
X-ray luminosity is directly related to the number (or mass) of stars in the
galaxy.  The NIR K-band luminosity is a more accurate measure of the galaxy
stellar mass and we prefer to use it for a proxy for the stellar mass instead
of the B-band luminosity.

For all 32 galaxies in our sample, we find that L$_{K}$ is slightly better
correlated with L$_{XP}$ (Pearson correlation coefficient r$=$0.93) than
is L$_B$ (r$=$0.90).  The best fit yields L$_K \propto$ L$_{XP}^{0.97}$.
When elliptical or spiral galaxies are considered separately, for both $B$ and
$K$, the correlation is quite strong, with $r \approx$ 0.9 (although
$r \approx$ 0.8 for $B$ and the spirals).

The FIR luminosity, which is an approximate 
measure of {\it current} star formation in late-type spirals and starburst
galaxies, is only 
correlated ($r >$ 0.50) with L$_{XP}$ for the Merger/Irr and 
Spiral galaxies.  
When all of the galaxies
are considered, there is no correlation between L$_{FIR}$ and L$_{XP}$
($r \approx$ 0.3).  For
ellipticals only, 
we find $r \approx$ 0.1.  The correlation is better ($r =$ 0.68) for the 
spiral galaxies.  We find the best correlation ($r =$ 0.93) for the
Merger/Irr galaxies, which, in general, 
have large SFR/M ratios (Table 2, columns 9 and 10).
This suggests
that a significant fraction of L$_{XP}$ in high SFR/M galaxies is due to 
current star formation, whereas in low-to-moderate SFR/M galaxies, much of
L$_{XP}$ could be due to the older population of X-ray sources.
The correlation between SFR indicators and the {\it total} X-ray
luminosity has been well noted in the literature 
(e.g. Helfand \& Moran 2001, Ranalli et al. 2003, Grimm et al. 2003).  
Here, we show specifically that the {\it point-source} X-ray luminosity
L$_{XP}$ is also well correlated with $SFR$ (and $M$).

Since some of the light from young, massive stars escapes directly from the
galaxy as UV radiation, we use
L$_{FIR+UV}$ (L$_{FIR} +$ L$_{FUV}$)
as a more accurate proxy for
the current SFR.  As with L$_{FIR}$, when all galaxies are 
considered, there is only a weak
correlation ($r \approx$ 0.5) between L$_{FIR+UV}$ and  L$_{XP}$.
We see a stronger correlation for the Merger/Irr and Spiral galaxies 
(r$=$0.94 and 0.69, respectively).
For ellipticals only, there is absolutely no correlation ($r \approx$ 0.01).

A careful inspection of the L$_{XP}$$-$L$_{K}$ 
scatter plot in Figure 1 (upper right) reveals
that the Merger/Irr and Spiral galaxies are systematically offset toward
larger L$_{XP}$, when compared to the Elliptical galaxies.  This merely
illustrates that there is a significant component of the 
point-source X-ray luminosity that is due
to current star formation and is not directly related to the galaxy
stellar mass.    We discuss this effect further in section 5.

\subsection{X-ray Point Source Luminosity Functions}

In Figures 2 and 3, we show the 
cumulative point source X-ray Luminosity Functions (XLFs)
N($>$L) for all of the points sources from each of the 32 datasets.  
We list the slope\footnote{%
We have taken special care to omit luminosity ranges on the lower end that
could be incomplete due to poor luminosity sensitivity within
the luminosity bin.  The exact ranges used are listed in the notes to Table 4.
We have also used a weighted least-squares technique to
compute $\gamma$, which gives much less weight to the highest luminosity bins,
compared to a simple least-squares technique.  
Our measurement of $\gamma$ may differ from other measurements
reported in the literature (e.g. Kilgard et al. 2002), especially if 
different luminosity ranges were used,
or the galaxy point-source lists are different.}
 $\gamma =$ $-$dlogN($>$L)/dlogL for each of the 
galaxies in Table 4.
For the Merger/Irr group, the mean and standard
deviation for $\gamma =$ 0.65$\pm$0.16, with the most prominent
outlier being the dwarf galaxy NGC~5253 with $\gamma \approx$ 0.9 (see Table 4).
The spirals have very similar slopes:
$\gamma =$ 0.79$\pm$0.24.
The luminosity functions of the elliptical galaxies, however, have steeper 
slopes: $\gamma =$ 1.41$\pm$0.38.
These differences in XLF slopes has been known for some time
(e.g., Primini et al. 1993, Kilgard et al. 2002, Eracleous et al. 2002).
The steeper slope for the ellipticals could be indicative of a different mode 
of X-ray binary formation, or, perhaps more likely, 
an older XRB population that is in a later stage of X-ray evolution
(e.g., see Grimm et al. 2002 and Wu 2001).

The total point source luminosity L$_{XP}$ can be theoretically defined as
$$L_{XP}^{theor.} = \int_{L_{min}}^{L_{max}} L n(L) dL \propto L_{max}^{1-\gamma} - L_{min}^{1-\gamma},     \eqno(2)$$
where, for simplicity, we have assumed $\gamma \ne$ 1.  Here
$n(L)$ is the differential luminosity function 
[$n(L) = N_0 \gamma L^{-(\gamma+1)}$ when N($>$L) $=$ N$_0$ L$^{-\gamma}$],
and L$_{min}$ and L$_{max}$ are the X-ray luminosities of the least and most
luminous X-ray point sources in the galaxy, respectively.
Since the slope of the cumulative XLF $\gamma < 1$ for the Merger/Irr and
spiral galaxies,
L$_{XP}$ is most sensitive to the upper limit L$_{max}$.
Although L$_{min}$ and L$_{max}$ are different for each galaxy, 
in general, the ratio L$_{max}$/L$_{min}$ $\gapprox$10$^2$
(see Figures 2 and 3), so that the L$_{max}$ term of the integral dominates
over the L$_{min}$ term by factors $\gapprox$5.0 and $\gapprox$2.6, for the
Merger/Irr and spiral galaxies, respectively.
This is quite interesting, since it implies that for galaxies with shallow 
XLFs, such as spiral
galaxies, the most luminous point sources, such as the ULXs, 
dominated L$_{XP}$.  Furthermore, for galaxies with several sources above 
10$^{38}$ erg~s$^{-1}$, observational measurements of L$_{XP}$ 
using very deep imaging data sensitive to very faint 
(e.g., $\gapprox$10$^{36}$ erg~s$^{-1}$) sources
should give approximately the same value for L$_{XP}$ as a shorter observation
sensitive to sources above  $\sim$10$^{37}$ erg~s$^{-1}$.  
Therefore, we argue
that the values we quote for
L$_{XP}$ in Table~3 are reliable for the spiral and Merger/Irr galaxies.

\subsection{X-ray Color-color diagrams \label{ccdiagrams}}

X-ray color-color (CC) diagrams were constructed using net counts in 
three X-ray bands, 
soft (S -- 0.3$-$1.0 keV), medium (M -- 1.0$-$2.0 keV), and 
hard (H -- 2.0$-$8.0 keV).  In our CC diagrams (Figure 4), we 
plot soft hardness
$MS = (M - S)/(M + S)$ against hard hardness
$HM = (H - M)/(H + M)$.
The Poisson uncertainty in $HM$ and $MS$ is $\lapprox$0.30 and $\lapprox$0.35,
respectively.\footnote{%
This estimate was determined empirically using the approximate Poisson error formula $\delta N \approx 1 + \sqrt{N + 0.75}$ (e.g., Gehrels 1986) for the raw counts.
For all of the sources with 20$-$100 net counts, the mean and standard deviation for
$\Delta HM$ and $\Delta MS$ are 0.30$\pm$0.21 and 0.34$\pm$0.57, respectively.
For sources with $>$100 net counts $\Delta HM =$ 0.10$\pm$0.14 and 
$\Delta MS =$ 0.11$\pm$0.14.}
Tabulated values of the net counts in each of the three bands, and
hardness ratios, are listed in Table A1.
In Figure 5, we show a grid of the
expected locations of X-ray sources, for
a simple power-law model with foreground absorption.
X-ray colors were simulated using the response of BI CCD~7.
No foreground absorption corresponds to the lowest
values of MS in the diagram, while increasing absorption pushes
the source up, and eventually to the right, when the 
medium band (1.0$-$2.0 keV) begins to be affected by absorption.
In Figure 5, we show the areas of the CC diagram that would be 
occupied by three sample types of point sources found in galaxies:
typical MW LMXBs, typical SMC HMXBs, which are much harder, and
(spiral galaxy) ULXs,
based on survey results from Church \& Balucinska-Church (2001),
Yokogawa (2002), Foschini et al. (2002), and Roberts et al. (2002).

We first examine the CC diagrams of the elliptical galaxies in Figure 4,
since they seem to be the least complex.  Most of the point sources in the
elliptical galaxies are consistent with the MW LMXBs (HM $\sim$ $-$0.5),
with little or no (N$_H <$ 3 $\times$ 10$^{21}$ cm$^{-2}$) absorption.
Some of the sources have harder X-ray colors (HM $\sim$0.25).
One would not expect HMXBs in elliptical galaxies, and we offer two
explanations.  The harder sources are predominantly the faintest
sources with a small number of total counts, so the uncertainties in
their X-ray colors is larger than the other sources.  Since the scatter
of the points in the CC diagram appears nearly symmetric, this seems
like the most likely explanation.  A second consideration is that the
MW LMXBs have lower X-ray luminosities (L$_X \sim$ 10$^{36}$$-$10$^{38}$
erg~s$^{-1}$) than the elliptical galaxy sources (L$_X \gapprox$ 10$^{38}$
erg~s$^{-1}$), so there could be an additional
smaller population of very-luminous
harder-spectrum LMXBs in elliptical galaxies.

We next turn to the spiral galaxies (see Figure 4).  There is noticeably more
scatter in the CC diagram for the spiral galaxy sources than for the elliptical
galaxy sources. The 
scatter is from the lower left to the upper right of the
diagram, as one would
expect for point sources with absorption columns ranging from negligible to 
$\sim$10$^{23}$ cm$^{-2}$.
Most of the sources in the spiral galaxies are not consistent with
the hard-spectrum SMC HMXBs, which generally have HM $\sim$ 0.25.  
Their colors are more consistent with an absorbed power-law spectrum
with $\Gamma \approx$ 1$-$2.
Even the most luminous sources (the ULXs, which are filled circles in 
Figure 4) in the
spiral and Merger/Irr galaxies have soft
($\Gamma \sim$ 1$-$2) spectra.
As mentioned in section 4.2, the most luminous X-ray point sources
dominate L$_{XP}$ in the spiral and Merger/Irr galaxies, and, for
most galaxies,
this corresponds to point sources with L$_X >$ 10$^{38}$ erg~s$^{-1}$
(see XLFs in Figure 4).  For convenience, in Figure 6, we 
show a CC diagram
for only the sources with L$_X >$ 10$^{38}$ erg~s$^{-1}$
in the spiral and Merger/Irr galaxies.
From this plot, it is quite obvious that 
hard-spectrum 
($\Gamma \approx$ 0.5$-$1) HMXBs are not dominating L$_{XP}$ in these galaxies.
The point-source luminosity
L$_{XP}$ in these two galaxy groups 
are ostensibly dominated by
young (pop~I) X-ray sources, so exactly what type of soft-spectrum sources are
those that are
plotted in Figure 6?
Obviously, the ULXs (filled circles) will dominate L$_{XP}$
when they are present; however, ULXs are generally present in only one of
every $\sim$5 disk galaxies (e.g. Sipior 2003).  
The sources with L$_X \approx$ 10$^{38}$$-$10$^{39}$ erg~s$^{-1}$ 
occupy essentially
the same regions of the CC diagram as the ULXs, i.e., 
{\it 
spectrally, there is no distinction between these two types of objects.}

We can examine the spectral properties of well-studied (i.e. the nearest)
luminous HMXBs to see if they are consistent with the hardness ratios of the
spiral galaxy sources.
In Table 5, we list all of the HMXBs with L$_X \ge$ 10$^{38}$ erg~s$^{-1}$,
as listed in the XRB catalog of Guseinov et al. (2000; 
v0.1, URL: {\sc www.xrbc.org}).  These nine objects 
are the most luminous HMXBs in the MW and Magellanic Clouds.  There
are three Be/transient sources, three accretion-powered X-ray pulsars,
two
black hole candidates, and Cyg~X-3, which is not well categorized.  We also
list the photon power law index $\Gamma$ from a simple absorbed power-law fit,
even for those XRBs with complex X-ray spectral models (see Table 5 footnote).

Two of the three HMXB transients have hard spectra ($\Gamma \lapprox$ 1), 
similar
to the SMC HMXBs plotted in Figure 5.  The transient EXO~2030$+$375 is soft
($\Gamma \sim$ 1.8) during outburst, but hardens to $\Gamma \sim$ 1 as it 
fades to $\sim$10$^{37}$ erg~s$^{-1}$ (e.g., Reig \& Coe 1999).
The accretion-powered NS HMXBs (SMC~X-1, LMC~X-4, and Cen~X-3) 
have high and low states that are similar to those of the BH HMXB Cyg~X-1 
(high/soft and low/hard), but they have much 
harder spectra in their high states ($\Gamma_{high} \sim$ 1, compared to
$\Gamma_{high} \sim$ 2.5  for Cyg~X-1).
The two black-hole candidate (BHC) HMXBs LMC~X-3 and LMC~X-1 have high 
states with $\Gamma \sim$ 2$-$3, similar to Cyg~X-1.
Cyg~X-3 has a relatively soft spectrum ($\Gamma \sim$ 2) in its high state, 
which hardens to $\Gamma \sim$ 1 in its low state, when the X-ray flux is
reduced by a factor of $\sim$2 (White \& Holt 1982).

Although a sample of nine luminous HMXBs is not a large sample, it is clear
that the different types of very-luminous HMXBs have fairly unique spectral
properties.
Based on the CC diagram in Figure 6,
the dominant X-ray sources in the spiral and Merger/Irr
groups are generally inconsistent with accretion-powered NS HMXBs, which 
have $\Gamma \lapprox$ 1.  They are more consistent with BHC HMXBs, such as
LMC~X-3
and LMC~X-1 in their soft state, with $\Gamma \sim$ 2$-$3.  
It is not clear what fraction of 
L$_{XP}$ from spiral and starbursting galaxies is from HMXB transients.
If we could extrapolate
from the MW and Magellanic Cloud HMXBs listed in Table 5, the
accretion-powered (non-transient) 
HMXBs should dominate the soft-spectrum X-ray luminosity,
since the X-ray luminosity of EXO~2030$+$375 is only $\sim$1/6 that of the 
combined X-ray luminosity of the ``soft'' accretion-powered HMXBs
LMC~X-3, LMC~X-1, and Cyg~X-3.

As we note in section 4.2, the pop~II X-ray sources have a steeper XLF,
and thus ``normal'' pop~I BH HMXBs and pop~I ULXs should dominate
L$_{XP}$ in the spiral and Merger/Irr galaxies.
It is important to note that many of the ULXs may, in fact, be pop~I BH HMXBs 
(e.g. King 2002), and, until ULXs are better understood, one should allow 
that there may be a substantial overlap between these two classes.
In other words, ULXs are not necessarily distinct from the many known types 
of XRBs.

We conclude that the total point-source
luminosity in spiral and starburst galaxies is {\it not} simply dominated by 
accretion powered, hard-spectrum NS HMXBs from the 
young stellar population.
It is dominated by the very-luminous, soft-spectrum accretion-powered BH HMXBs,
and soft-spectrum ULXs.  These sources seem to 
have knowledge of (and are correlated with) both the current SFR, as measured
by L$_{FIR}$ and L$_{FIR+UV}$, and the mass of the galaxy, as measured by
L$_B$ and L$_{K}$.  
Since {\it the exact nature of ULXs are not well known}
and they will dominate L$_{XP}$, if present, 
it is premature to say that the X-ray emission from 
the young stellar population is dominated by ``high-mass X-ray binaries'' (e.g. 
Grimm et al. 2003).

By inspecting the locations of the ULXs in the CC diagrams in Figure 4, one 
notices that, whereas the ULXs in the elliptical galaxies are clustered near 
HM $\sim$ $-$0.4, the ULXs in the Merg/Irr and Spiral groups are slightly
harder in HM.  For example, there are no
ULXs with HM $<$ $-$0.4 in the spiral group.  This could possibly be a result
of obscuration by the gas in spirals, but this would require the ULXs to be
highly obscured.  As our simulations in Figure 5 (top left plot) show, 
absorption columns $\gapprox$10$^{22}$ cm$^{-2}$ are required to increase
HM by a factor $\gapprox$0.4.
Alternatively, this could be a 
manifestation of two physically different types of ULXs (e.g., King 2002).

\section{Discussion}

\subsection{Summary of Results}

In the previous section, we showed that the total {\it point source}
X-ray luminosity
L$_{XP}$ is indeed correlated with properties of the host galaxy that are 
associated with the mass and SFR in the galaxy.  Based on the XRB populations
in our Galaxy, one might therefore suppose that the correlations merely
represent an extrapolation of the total X-ray luminosity from the LMXB
and HMXB populations, scaled to the galaxy mass and SFR 
(e.g., Grimm et al. 2002).
Using X-ray color-color diagrams, we showed
that the SFR component of L$_{XP}$ is dominated by soft-spectrum BH HMXBs and
ULXs, not hard-spectrum NS HMXBs.  
Since ULXs are not present in our Galaxy, X-ray studies of 
external galaxies are necessary to understand their effect on X-ray 
point source populations.
Since we can actually 
measure the correlation between L$_{XP}$ and either L$_K$ or
L$_{FIR+UV}$, both the mass and SFR components of L$_{XP}$ are significantly
strong.
Since the slope $\gamma$ of the XLFs is shallow for the spiral galaxies,
the most luminous point sources 
(i.e., BH HMXBs and ULXs with L$_X \gapprox$ 10$^{38}$ 
erg~s$^{-1}$) dominate L$_{XP}$.  
We emphasize that the ULX distinction is purely based on X-ray luminosity
and does not necessarily represent a unique, or homogeneous class of objects.
For the spiral and Merger/Irr galaxies, we find no particular distinction 
between the X-ray colors of the ULXs and the less luminous point sources
with L$_X \approx$ 10$^{38}$$-$10$^{39}$ erg~s$^{-1}$.
As our CC diagrams show, the X-ray colors of the point sources are generally
consistent with LMXBs or ULXs, not hard-spectrum NS HMXBs, such as accretion-powered
X-ray pulsars.
This is also true of the most 
luminous point sources (e.g., the ULXs).

\subsection{Multi-variate Correlations Between L$_{XP}$ and 
Host Galaxy Luminosities}

We would like to deconvolve L$_{XP}$ into two components, one from the 
older stellar population (including LMXBs), and another from the
younger stellar population (including HMXBs).  We assume the old and
young components are directly related to the mass and SFR of the galaxy,
respectively, and initially explore a simple relationship between L$_{XP}$
and the host galaxy K and FIR+UV luminosities, which are reasonable proxies
for the mass and SFR.

We use the simple linear relationship
$$L_{XP} = A L_{K} + B L_{FIR+UV},      \eqno(3)$$
where $A$ and $B$ are dimensionless constants.
We have used $\chi^2$ fitting to estimate A and B and their uncertainties.\footnote{%
Errors for L$_K$ and L$_{FIR}$ were estimated from the errors in the flux measurements
(see references in notes to Table 2).  We estimated a 30\% error for the FUV fluxes,
due to uncertainties in fitting SEDs to the data (see section 3.3).
We also experimented with other fitting methods 
(principal component analysis, simple least-squares, 
and robust weighted least-squares) and arrive at the same 
$A$ and $B$ values, within the $\chi^2$ uncertainties.
}
The best fit values of $A$ and $B$ are listed in Table 6.
Since star formation is negligible in the elliptical galaxies, 
we can also estimate $A$ directly
from L$_{XP}$ by ignoring the SFR term.  
If we omit the peculiar elliptical galaxy
NGC~5128 (Cen~A), which has ongoing star-formation, $\chi^2-$fitting yields
$A =$ 1.26$^{+0.30}_{-0.31}$ $\times$ 10$^{-4}$.
This value for ellipticals is $\sim$20\% larger than that found from
multi-variate $\chi^2$ fitting for samples involving Merger/Irr and 
Spiral galaxies (see Table 6), although the two values are consistent
within the uncertainties.
Since AGN emission in the K, FIR and FUV bands has not been subtracted from
any of the host galaxy luminosities, we computed $A$ and $B$ for two
subsamples of galaxies without AGNs, and arrive at a slightly smaller
(larger)
value for $A$ ($B$) than if AGNs are included.
If the elliptical galaxies are also omitted from the sample, 
the best fit value
of $A$ is also lower.
We then suggest that the value of $A$ for early-type and elliptical galaxies
is inherently different from that in late-type galaxies.

The fact that $A$ and $B$ are the same order of magnitude, and that L$_K$
and L$_{FIR+UV}$ are also the same order of magnitude, supports our
earlier statement that both the old (mass) and young (SFR) component
can contribute significantly to L$_{XP}$.

We conclude, based on our best fit to the Merger/Irr plus spiral 
sample without AGNs, that the
the point-source X-ray luminosity L$_{XP}$ in Merger/Irr and spiral 
galaxies can be reliably estimated from:
$$L_{XP} = (0.9\pm0.1) \times 10^{-4} L_K + 
  (0.5\pm0.1) \times 10^{-4} L_{FIR+UV},            \eqno(4)$$
and that L$_{XP}$ for elliptical galaxies can be estimated from 
$$L_{XP} = (1.3\pm0.3) \times 10^{-4} L_K.           \eqno(5)$$

The robustness of this formula for the Merger/Irr and spiral
galaxies can be seen in Figure 7 (top), where we plot
L$_{XP}$ against the estimated value using equation 4.
The Merger/Irr galaxies and Spiral galaxies are both in good agreement,
although L$_{XP}$ for the elliptical galaxies is under-estimated.
As we argue in the next section, 
the larger (M/L)$_K$ ratio for ellipticals (see Table 7) should likely cause
the $A$ value for
ellipticals and spirals to be inherently different.  

We note that the reduced-$\chi^2$, $\chi^2_\nu$, for all of the fits 
(see Table 6) is highly 
sensitive to the estimated errors in the luminosity uncertainties, which
are not well defined.  
Therefore, we do not use $\chi^2_\nu$ as an absolute measure of the
goodness of fit.
We also note that the ranges of $A$ for spirals and ellipticals overlap,
and we cannot rigorously show that the values of $A$ are different for 
ellipticals and spirals, although, as we show in the next section, the effect
from including an M/L correction is very noticeable.

\subsection{Multi-variate Correlations between L$_{XP}$ and M and SFR}

We next write L$_{XP}$ in terms of the mass M and SFR of the galaxy
$$L_{XP} = \alpha M + \beta SFR,     \eqno(6)$$
where L$_{XP}$, M, and SFR have units erg~s$^{-1}$, 
M$_\odot$, and M$\odot$~yr$^{-1}$, respectively,
and $\alpha$ and $\beta$ are constants with the appropriate units.

We use the optical/NIR B$-$K color and the K-band luminosity (Table 2)
to estimate the mass of the sample galaxies, and L$_{FIR+UV}$ to estimate
the SFR.  Bell \& de~Jong (2001) list correlation coefficients for M/L
ratios as a function of optical and NIR colors.  We use the coefficients
for their formation model with bursts.  The SFR can be estimated from
L$_{FIR+UV}$ if we assume L$_{FIR+UV}$ is a ``corrected'' IRAS FIR
luminosity.  We use a proportionality constant of
5.7 $\times$ 10$^{-44}$ M$_\odot$~yr$^{-1}$~(erg~s$^{-1}$)$^{-1}$,
which includes a bolometric correction of 1.4 from the IRAS FIR band
(Meurer et al. 1999).  The resulting values for SFR, (M/L)$_K$ and $M$
are listed in Table 7.

Again, we find $\alpha$ and $\beta$ and their uncertainties using
minimum-$\chi^2$ techniques.  The uncertainties in M and SFR were computed
from the uncertainties in all of the luminosities (or magnitudes) used
in computing M and SFR.
In Table 6, we list the resulting values for the Merger/Irr plus spirals
sample (omitting AGNs).  We find that the mass-component constant
$\alpha$ is slightly different for the elliptical and spiral samples, 
but this time $\alpha$ is smaller for the ellipticals, although
both $\alpha$ values are consistent within their uncertainties.  
As can be seen in Figure 7 (bottom), with the M/L correction applied,
the simulated value of L$_{XP}$ now agrees better with the 
observed value.
The simulated value of L$_{XP}$ for the elliptical galaxies is now much closer
to the observed value (compare top and bottom plots in Figure 7).
This indicates
that $\alpha =$ 1.25 $\times$ 10$^{29}$ erg~s$^{-1}$~M$_\odot^{-1}$ is a 
{\it universal constant} for estimating the point-source X-ray luminosity
from the older stellar population.

Thus, we find the following empirical relationship to hold for 
high-SFR/M Merger/Irr galaxies, spiral galaxies, and
low-SFR/M elliptical galaxies:
$$L_{XP}(erg~s^{-1})  = (1.3\pm0.2) \times 10^{29} M(M_\odot) + 
  (0.7\pm0.2) \times 10^{39} SFR(M_\odot~yr^{-1})      \eqno(7).$$

As noted in the caption to Figure 7, the observed values for L$_{XP}$ for the
two starburst galaxies NGC~4449 and NGC~4038/9 are in excess of the values
predicted by Equations (4) and (7), by factors of $\approx$4 and $\approx$5,
respectively.  Since L$_{XP}$ is dominated by the $SFR$ component for starburst
galaxies, an error in the $SFR$ by $\sim$4$-$5 would be needed to ``correct''
the problem.  This is unlikely.  Poisson uncertainty in the number of
high-luminosity sources (which dominate L$_{XP}$) could possible explain
the excess, but it is also possible that these two starburst galaxies have 
significantly high efficiency for forming luminous BH~HMXBs and ULXs.  It
is interesting that NGC~4038/9 is a merging system and NGC~4449 has evidence
for two physical counter-rotating systems within the galaxy (Sabbadin
et al. 1984).  More extensive tests would be needed to determine
if merger activity is really responsible for the abnormally high values of
L$_{XP}$ in these two galaxies.

\subsection{The Linearity of L$_{XP}$, $M$, and $SFR$}

In order to test the linear form we have assumed
in Equation (7), we performed a number of tests.
A least-squares fit between $logL_{XP}$ and $logM$ for 
the eight elliptical galaxies in our sample with no star formation (NGC~5128 
omitted) implies a best fit for $L_{XP} \propto M^{0.99}$ (r $=$ 0.89).
A more rigorous test from a 
multi-variate fit to the sample of 18 Merg/Irr and spiral galaxies without
AGN gives the best fit for the mass component of L$_{XP} \propto M^{1.01}$
(See Figure 9).  Therefore, we can argue with good confidence that the
pop~II component of L$_{XP}$ is essentially linear in $M$.
Similarly, a
multi-variate fit to $L_{XP} = \alpha M + \beta SFR^{p_{SFR}}$
gives a best fit for p$_{SFR} =$ 1.06, with a 90\% confidence range of
0.85$-$1.25 (for three free parameters: $\Delta\chi^2 <$ 6.25; see Figure 9).
We arrive at the a very similar result ($p_{SFR} =$ 1.09) if we use only
the five Merger/Irr galaxies, which have a very weak mass component, and fit 
$log($L$_{XP})$ against $log(SFR)$ ($r =$ 0.94).  Thus, 
within the uncertainties,
{\it we find L$_{XP}$ to be linear in both the mass and the star-formation 
rate.}

\subsection{Nature of the ULXs}

As mentioned previously, we assume the mass component of L$_{XP}$ is from point
sources from an older stellar population (including LMXBs), and the SFR 
component is from point sources from a younger stellar population
(including HMXBs).  It is not known if ULXs, which can dominate L$_{XP}$ (section 4.2),
are predominantly associated with older or younger stellar population, or are
a largely heterogeneous group (e.g., King 2002).  Thus, 
they could contribute significantly in
either component.  We know that ULXs exist in elliptical galaxies 
(e.g. Colbert \& Ptak 2002).
If LMXBs are formed in globular clusters (e.g. Kundu et al. 2003), pop~II
ULXs might also be formed in the same environment.
On the other hand, in the high-SFR
galaxy pair NGC 4038/9 (The ``Antennae''), L$_{FIR}$ is very high and an 
extraordinarily large number of ULXs are found 
(Fabbiano, Zezas \& Murray 2001), suggesting that
the ULXs in the Antennae are associated with the young stellar population.
So what about ULXs in the normal spirals, in which old and young stellar
populations (ie the K and FIR+UV luminosities) can be comparable?

In Figure 8, we plot the fraction of L$_{XP}$ from the younger stellar 
population against galaxy morphological type.  As expected, in late-type
galaxies with high SFR/M ratios, the SFR term dominates, and in early-type
galaxies, the mass term dominates.  This plot also shows that the older
population contributes $\approx$20$-$90\% of L$_{XP}$ in Merger/Irr and
spiral galaxies.  As described in section 4.2, since the slope of the XLF
for these groups is shallow ($\gamma \lapprox$ 1), 
the most luminous
X-ray point sources dominate L$_{XP}$. 
This would imply that,
if the old and young components of the 
XLF have exactly the same shape, 
$\sim$20$-$90 percent of the ULXs are
from the older population.  However, this is not likely to be the case since
the XLFs for older point sources in the elliptical galaxies
is much steeper than the XLFs for old+young point sources in the spirals,
and many of the ULXs are from the younger population.  However, it does suggest
that type~II (pop~II) ULXs could have significant numbers in spiral galaxies.
Since we now know that L$_{XP}$ from the older stellar population is 
proportional to the galaxy mass $M$,  we can estimate the fraction of
type~II ULXs in spirals by scaling the number found in ellipticals
by the galaxy mass $M$.  If we exclude NGC~5128 since it has ongoing
star formation, the total mass of the remaining eight ellipticals is
1.06 $\times$ 10$^{12}$ M$_\odot$, and eleven ULXs are found.  This
implies an expected 1.0 $\times$ 10$^{-11}$ ULXs~M$_\odot^{-1}$, with
a Poisson error of 30\%.  The total mass of all of the Merg/Irr and 
spiral galaxies is 6.6 $\times$ 10$^{11}$ M$_\odot$, so we expect 
$\approx$5$-$9 pop~II ULXs to be found.  A total of 32 ULXs were detected,
implying that, on average, {\it $\sim$15$-$25\% of the ULXs in Merger/Irr and
spiral galaxies
are pop~II ULXs.}  
The same calculation for the Merger/Irr and spiral galaxy samples alone
yield rates of $\sim$5$-$10\% and $\sim$25$-$50\% of pop~II ULXs for
the ``high-SFR/M'' Merger/Irr galaxies, and normal spirals, respectively.

\section{Summary and Conclusions}

We have used archival and proprietary Chandra ACIS data for
32 spiral, elliptical, and Merger/Irr (starburst) galaxies to study
X-ray point source populations, and how they depend on properties of
the host galaxy.  A total of 1441 X-ray point sources are analyzed here, 
using the CPU-intensive XASSIST data reduction scripts 
(Ptak \& Griffiths 2003) to rigorously verify the
validity of the point sources.

The total {\it point-source} X-ray (0.3$-$8.0 keV) luminosity L$_{XP}$ is well
correlated with the B-band, K-band, and FIR+UV luminosities of spiral host
galaxies, and is well correlated with the B-band and K-band luminosities
for elliptical galaxies.  This has been known for some time, and it
suggests an intimate connection between
L$_{XP}$ and both the old and young stellar populations, for which K
and FIR+UV luminosities are reasonable proxies for the galaxy mass $M$ and 
star-formation rate $SFR$.  

We derive proportionality constants
$\alpha =$ 1.3 $\times$ 10$^{29}$ erg~s$^{-1}$~M$_\odot^{-1}$ and 
$\beta =$ 0.7 $\times$ 10$^{39}$ erg~s$^{-1}$~(M$_\odot$~yr$^{-1}$)$^{-1}$,
which can be used to estimate the old and young components from $M$ and 
$SFR$, respectively.  

The cumulative X-ray luminosity functions for the
point sources have significantly different slopes.  For the spiral and
starburst galaxies, $\gamma \approx$0.6$-$0.8, and for the elliptical galaxies,
$\gamma \approx$1.4.  This implies that
{\it the most luminous point sources -- those with
L$_X \gapprox$ 10$^{38}$ erg~s$^{-1}$ --- 
dominate L$_{XP}$ for the spiral and
starburst galaxies.}
We find that L$_{XP}$ depends {\it linearly} (within uncertainties) on both
$M$ and $SFR$, for our sample galaxies ($M \lapprox$ 10$^{11}$ M$_\odot$ and
$SFR \lapprox$ 10 M$_\odot$~yr$^{-1}$).

Most of the point sources have X-ray colors that are consistent with 
soft-spectrum ($\Gamma \sim$ 1$-$2)
low-mass X-ray binaries (XRBs), accretion-powered black-hole high-mass 
(BH HMXBs), or Ultra-Luminous X-ray sources (ULXs a.k.a. IXOs).
We rule out hard-spectrum neutron-star HMXBs (e.g. accretion-powered X-ray
pulsars)
as contributing much to L$_{XP}$.  
Thus, for spirals, L$_{XP}$ is dominated by ULXs and BH HMXBs.
We find no discernible difference between the X-ray colors of ULXs (L$_X \ge$
10$^{39}$ erg~s$^{-1}$) in spiral galaxies and point sources with 
L$_X \approx$ 10$^{38}$$-$10$^{39}$ erg~s$^{-1}$, suggesting ULXs in spiral
galaxies could be predominantly BH HMXBs.  This would not be surprising,
since ULXs are defined purely by their X-ray luminosity, and not by their
accretion mode, or type of companion star.  More work on ULXs is needed to
test whether most are simply a special mode of a ``normal'' stellar-mass
BH XRB (e.g. King 2002), or exotic accreting intermediate-mass BH
systems (e.g. Colbert \& Mushotzky 1999).

We estimate that $\gapprox$20\% of all ULXs found
in spirals originate from the older (pop~II) stellar populations, indicating
that many of the ULXs that have been found in spiral galaxies are in fact
pop~II ULXs, like those in elliptical galaxies.

\acknowledgments

We thank the anonymous referee for many helpful suggestions that helped to
improve the quality of the paper.
We are grateful to T. Jarrett for providing 2MASS K$_s$ magnitudes from the
Large Galaxy Atlas before publication.  We thank A. Zezas, T. Roberts,
T. Yaqoob and A. Prestwich
for helpful discussions, and T. Budavari for expert help with SED fitting.
EJMC acknowledges support from NASA grant NAG 5-11670.

\appendix

\section{Tabulation of X-ray Point-source Properties}

As mentioned in section 1, uniform data processing of all of the datasets is important
for large samples of galaxies.  We list in Table \ref{appendixtable} some of the
X-ray properties for each of the 1441 point sources in the 32 galaxies in our sample.

\clearpage

\noindent
{\bf Figures are not included here.  Please download full manuscript from
the following website for version with figures.  Sorry.  This URL is also 
listed in the notes to the astro-ph abstract, so you may be able to just
click on it there.

http://www.pha.jhu.edu/$\sim$colbert/chps\_accepted.ps
}

\clearpage

\clearpage

\begin{deluxetable}{llrccr}
\tabletypesize{\scriptsize}
\tablecaption{Observational Data \label{tab1}}
\tablewidth{0pt}
\tablehead{
\colhead{Galaxy} & \colhead{Other} & \colhead{ObsID} & \colhead{CCDs} & \colhead{Exp. Time} & 
\colhead{N} \\
\colhead{Name} & \colhead{Name} & 
\colhead{} & \colhead{} & \colhead{(ks)} & \colhead{} \\     
\colhead{(1)} & \colhead{(2)} & \colhead{(3)} & 
\colhead{(4)} & \colhead{(5)} & \colhead{(6)} \\
}
\startdata
NGC~1569   &          &  782 & 7         & 85.1                     & 12 \\ 
NGC~3034   & M82      &  361 & 0/1/2/3   & 33.2/32.8/33.1/33.0      & 27 \\
NGC~4038/9 & Antennae &  315 & 6/7       & 72.1/70.2                & 65 \\
NGC~4449   &          & 2031 & 7         & 26.2                     & 22 \\
NGC~5253   &          & 2032 & 7         & 46.7                     & 10 \\
\tableline
NGC~253    &          &  969 & 2/3/6/7   & 13.9/13.9/13.9/11.8      & 49 \\
NGC~628    & M74      & 2057 & 6/7       & 46.2/44.7                & 60 \\
NGC~1291   &          & 2059 & 6/7       & 11.2/12.0                & 56 \\
NGC~2681   &          & 2061 & 7         & 78.3                     & 17 \\
NGC~3079   &          & 2038 & 7         & 26.3                     & 17 \\
NGC~3184   &          &  804 & 7         & 37.1                     & 45 \\
NGC~3628   &          & 2039 & 7         & 54.8                     & 32 \\
NGC~4244   &          &  942 & 7         & 48.1                     &  3  \\
NGC~4258   & M106     &  350 & 6/7       & 14.0/13.9                & 26 \\
NGC~4314   &          & 2062 & 7         & 8.3                      & 12 \\
NGC~4579   & M58      &  807 & 7         & 32.8                     & 8 \\
NGC~4631   &          &  797 & 6/7       & 59.1/57.6                & 25 \\
NGC~4736   & M94      &  808 & 7         & 47.3                     & 28 \\
NGC~4945   &          &  864 & 6/7       & 35.2/5.1                 & 36 \\
NGC~5194/5 & M51a/b   &  354 & 6/7       & 14.8                     & 52 \\
NGC~5236   & M83      &  793 & 6/7       & 50.9/48.2                & 84 \\
NGC~5457   & M101     &  934 & 2/3/5/6/7 & 98.1/98.1/96.3/98.2/95.6 & 140 \\
NGC~6503   &          &  872 & 7         & 9.1                      &  7 \\
\tableline
NGC~1395   &          &  799 & 3         & 22.2                     & 30 \\
NGC~1399   &          &  319 & 7         & 55.5                     & 100 \\
NGC~3379   & M105     & 1587 & 7         & 31.0                     & 35 \\
NGC~4374   & M84      &  803 & 7         & 28.3                     & 28 \\
NGC~4472   & M49      &  321 & 6/7       & 39.2/33.8                & 78 \\
NGC~4636   &          &  323 & 7         & 45.2                     & 37 \\
NGC~4649   & M60      &  785 & 6/7       & 36.7/22.3                & 98 \\
NGC~4697   &          &  784 & 7         & 39.0                     & 69 \\
NGC~5128   & Cen~A    &  316 & 0/1/2/3   & 35.5/35.3/35.4/35.5      & 133 \\
\enddata

\tablenotetext{~}{ Notes on table columns: 
(1,2) Galaxy Names; (3) Chandra Observation ID; 
(4) ACIS CCD numbers in galaxy FOV for which point sources were detected;
(5) Exposure time for CCDs listed in column 4; 
(6) Number of X-ray point sources detected within galaxy FOV (see section 3 for 
details). 
}

\end{deluxetable}

\clearpage

\begin{deluxetable}{lccccccccc}
\tabletypesize{\scriptsize}
\tablecaption{Galaxy Properties \label{tab2}}
\tablewidth{0pt}
\tablehead{
\colhead{Galaxy} & 
  \colhead{Distance} & \colhead{Morph.} & \colhead{AGN} & 
  \colhead{L$_B$} & \colhead{L$_{K}$} & 
  \colhead{L$_{FIR}$} & \colhead{L$_{UV}$} & 
  \colhead{${{L_{FIR}}\over{L_B}}$} & \colhead{${{L_{FIR+UV}}\over{L_{K}}}$} \\
\colhead{Name} & 
\colhead{(Mpc)} & \colhead{Type} & \colhead{Type} & 
\colhead{($log$ erg/s)} &
\colhead{($log$ erg/s)} &
\colhead{($log$ erg/s)} &
\colhead{($log$ erg/s)} &
  \colhead{} & \colhead{} \\
\colhead{(1)} & \colhead{(2)} & \colhead{(3)} & \colhead{(4)} & 
\colhead{(5)} & \colhead{(6)} & \colhead{(7)} & \colhead{(8)} &
\colhead{(9)} & \colhead{(10)} \\
}
\startdata
NGC 1569   &  1.6 & Sm            &       & 42.2 & 41.3 & 41.8 & 40.3 &  0.4 &  3.6 \\ 
NGC 3034   &  5.2 & I0            &       & 43.4 & 43.6 & 44.2 & 41.7 &  6.0 &  4.0 \\ 
NGC 4038/9 & 21.7 & Sc/Sc (tides) &       & 44.3 & 43.8 & 44.1 & 43.4 &  0.7 &  2.2 \\ 
NGC 4449   &  3.0 & Sm            &       & 42.5 & 42.0 & 42.4 & 42.1 &  0.7 &  3.6 \\ 
NGC 5253   &  3.2 & I0            &       & 42.4 & 41.7 & 42.2 & 41.9 &  0.7 &  5.5 \\ 
\tableline
NGC 253    &  3.0 & Sc            &       & 43.7 & 43.5 & 43.6 & 41.8 &  0.9 &  1.4 \\ 
NGC 628    &  9.7 & Sc            &       & 43.6 & 43.1 & 43.2 & 42.8 &  0.4 &  1.7 \\ 
NGC 1291   &  8.6 & SBa           &       & 43.7 & 43.6 & 42.1 & 42.1 &  0.02 & 0.06 \\ 
NGC 2681   & 13.3 & Sa            &       & 43.4 & 43.3 & 42.9 & 41.9 &  0.3 &  0.5 \\ 
NGC 3079   & 15.0 & Sc pec:       & S2    & 43.7 & 43.5 & 43.8 & 42.2 &  1.3 &  2.4 \\ 
NGC 3184   &  8.7 & Sc            &       & 43.3 & 42.9 & 42.8 & 42.6 &  0.3 &  1.3 \\ 
NGC 3628   &  7.7 & Sbc           &       & 43.6 & 43.3 & 43.3 & 42.4 &  0.5 &  1.1 \\ 
NGC 4244   &  3.1 & Scd           &       & 42.8 & 41.8 & 41.6 & 41.6 &  0.06 & 1.3 \\ 
NGC 4258   &  6.8 & Sb            & S1.9  & 43.8 & 43.5 & 43.0 & 42.4 &  0.2 &  0.4 \\ 
NGC 4314   &  9.7 & SBa           &       & 43.0 & 43.0 & 42.4 & 41.6 &  0.2 &  0.3 \\ 
NGC 4579   & 20.3 & Sab           & S1.9/L1.9 & 44.0 & 44.0 & 43.3 & 42.7 &  0.2 &  0.2 \\ 
NGC 4631   &  6.9 & Sc            &       & 43.8 & 43.1 & 43.3 & 42.9 &  0.3 &  2.3 \\ 
NGC 4736   &  4.3 & RSab          &       & 43.3 & 43.2 & 42.9 & 42.3 &  0.4 &  0.6 \\ 
NGC 4945   &  5.2 & Sc            & AGN   & 44.0 & 43.7 & 43.8 & 41.7 &  0.6 &  1.4 \\ 
NGC 5194/5 &  7.7 & Sbc/SB        & S2    & 43.9 & 43.8 & 43.3 & 43.1 &  0.2 &  0.5 \\ 
NGC 5236   &  4.7 & SBc           &       & 43.7 & 43.5 & 43.2 & 42.5 &  0.4 &  0.7 \\ 
NGC 5457   &  5.4 & Sc            &       & 43.7 & 43.1 & 43.3 & 43.4 &  0.4 &  3.2 \\ 
NGC 6503   &  6.1 & Sc            &       & 43.1 & 42.7 & 42.4 & 42.0 &  0.2 &  0.8 \\ 
\tableline
NGC 1395   & 22.7 & E2            &       & 44.0 & 43.9 & 41.6    & 42.3 & 0.003 & 0.03 \\               
NGC 1399   & 19.3 & E1            &       & 43.9 & 44.0 & 41.3\tablenotemark{a} & 42.5 & 0.002  & 0.03 \\
NGC 3379   &  8.1 & E0            &       & 43.3 & 43.3 & $<$40.3 & 41.7 & $<$0.001 & $<$0.02 \\        
NGC 4374   & 16.8 & E1            &       & 44.0 & 44.0 & 42.0    & 42.4 & 0.01  & 0.04 \\               
NGC 4472   & 16.8 & E1/S0         & S2::  & 44.3 & 44.3 & $<$41.1 & 42.8 & $<$0.0006 & $<$0.03 \\        
NGC 4636   & 14.6 & E0/S0         &       & 43.7 & 43.7 & 41.2\tablenotemark{a} & 42.4 & 0.003 & 0.05 \\ 
NGC 4649   & 14.9 & S0            &       & 44.0 & 44.1 & 42.0    & 42.5 & 0.01 & 0.03 \\               
NGC 4697   & 16.5 & E6            &       & 43.9 & 43.9 & 42.0    & 42.3 & 0.01 & 0.04 \\               
NGC 5128   &  4.9 & S0+S pec      & AGN   & 44.0 & 43.8 & 43.4    & 42.5 & 0.3  & 0.4 \\               
\enddata

\tablenotetext{~}{Notes on columns:
(1) Galaxy name;
(2) Distance to galaxy (see section 2);
(3) Galaxy morphological type, from the RSA galaxy catalog (Sandage \& Tammann 1981);
(4) AGN type, as listed in Ho et al. (1997).  We have additionally classified NGC~4945 and NGC~5128 as
    AGN, due to hard X-ray spectra properties, and nuclear jet, respectively.  Blank entries denote
    LINER 1.9/2, transition, and H~II classifications from Ho et al., or no evidence for AGN activity
    in the NED notes for that galaxy.
(5) Galaxy B-band luminosity $\nu$$L\nu$, calculated using B$_T^0$ (or m$_B^0$ for NGC 4038/9)
    magnitudes from RC3;
(6) Galaxy K$_s$-band luminosity $\nu$$L\nu$, calculated from 20 mag/sq. arcsec isophotal magnitudes
    from the 2MASS Large Galaxy Atlas (Jarrett et al. 2003);
(7) Galaxy FIR luminosities, calculated from IRAS fluxes using the method of 
  Fullmer \& Lonsdale (1989).  IRAS fluxes were taken preferentially
  from Moshir et al. (1997, as listed in NED), Soifer et al. (1989),
  Rice et al. (1988), Knapp (1994), Knapp et al. (1989), and 
  Thronson et al. (1987);
(8) Galaxy far-ultraviolet luminosity (see section 3.3); (9) Ratio of
FIR to B-band luminosity; (10) Ratio of FIR$+$UV to K$_s$ luminosity.  }

\tablenotetext{a}{%
Only upper limits were available for the 60$\mu$m IRAS flux for 
NGC~1399 and the 100$\mu$m flux for NGC~4636 (Knapp 1994).
The L$_{FIR}$ value listed here was calculated using 50\% of the 
quoted upper limit.  The corresponding uncertainties in L$_{FIR}$ 
due to this estimate are 0.05 and 0.08 dex for 
NGC~1399 and NGC~4636, respectively.
}

\end{deluxetable}

\clearpage
\begin{deluxetable}{lcccclclc}
\tabletypesize{\scriptsize}
\tablecaption{Galaxy X-ray Properties\label{tab2b}}
\tablewidth{0pt}
\tablehead{
\colhead{Galaxy} & \colhead{N$_H$(Gal)} & \colhead{L$_{X}$(limit)} & \colhead{L$_{XP}$} & 
\multicolumn{2}{c}{fraction $\ge$ 10$^{38}$} & \multicolumn{2}{c}{fraction $\ge$ 10$^{39}$} &
\colhead{ $log({{L_{XP}} \over {L_{K}}})$ } \\
\colhead{Name} & \colhead{($log$ cm$^{-2}$)} & 
\colhead{($log$ erg/s)} & \colhead{($log$ erg/s)} & 
\colhead{L$_{XP}$} & \colhead{N} & \colhead{L$_{XP}$} & \colhead{N} &
\colhead{} \\
\colhead{(1)} & \colhead{(2)} & \colhead{(3)} & \colhead{(4)} &
\colhead{(5)} & \colhead{(6)} & \colhead{(7)} & \colhead{(8)} & 
\colhead{(9)} \\
}
\startdata
NGC 1569   & 21.4 & 35.4 & 37.9 & 0.00 &  0.00(0)  & 0.00 & 0.00(0)  & $-$3.4 \\
NGC 3034   & 20.6 & 36.9 & 40.0 & 0.97 &  0.56(15) & 0.61 & 0.11(3)  & $-$3.6 \\
NGC 4038/9 & 20.6 & 37.7 & 40.8 & 0.99 &  0.82(53) & 0.77 & 0.22(14) & $-$3.0 \\
NGC 4449   & 20.1 & 36.3 & 39.0 & 0.66 &  0.14(3)  & 0.00 & 0.00(0)  & $-$3.0 \\
NGC 5253   & 20.6 & 36.2 & 38.0 & 0.00 &  0.00(0)  & 0.00 & 0.00(0)  & $-$3.6 \\
\tableline 
NGC 253    & 20.2 & 36.7 & 39.4 & 0.65 &  0.12(6)  & 0.00 & 0.00(0)  & $-$4.1 \\
NGC 628    & 20.7 & 37.2 & 39.6 & 0.54 &  0.15(9)  & 0.00 & 0.00(0)  & $-$3.6 \\
NGC 1291   & 20.3 & 37.6 & 39.7 & 0.65 &  0.23(13) & 0.00 & 0.00(0)  & $-$3.8 \\
NGC 2681   & 20.4 & 37.2 & 39.7 & 0.93 &  0.53(9)  & 0.59 & 0.12(2)  & $-$3.6 \\
NGC 3079   & 19.9 & 37.7 & 39.6 & 0.84 &  0.41(7)  & 0.43 & 0.06(1)  & $-$3.8 \\
NGC 3184   & 20.0 & 37.1 & 39.4 & 0.44 &  0.07(3)  & 0.00 & 0.00(0)  & $-$3.5 \\
NGC 3628   & 20.3 & 36.8 & 39.5 & 0.71 &  0.09(3)  & 0.63 & 0.03(1)  & $-$3.9 \\
NGC 4244   & 20.2 & 36.1 & 37.6 & 0.00 &  0.00(0)  & 0.00 & 0.00(0)  & $-$4.2 \\
NGC 4258   & 20.1 & 37.3 & 39.6 & 0.83 &  0.42(11) & 0.00 & 0.00(0)  & $-$3.8 \\
NGC 4314   & 20.3 & 37.9 & 39.0 & 0.66 &  0.33(4)  & 0.00 & 0.00(0)  & $-$3.9 \\
NGC 4579   & 20.4 & 37.9 & 40.2 & 1.00 &  1.00(8)  & 0.87 & 0.25(2)  & $-$3.8 \\
NGC 4631   & 20.1 & 36.7 & 39.5 & 0.86 &  0.20(5)  & 0.53 & 0.04(1)  & $-$3.6 \\
NGC 4736   & 20.2 & 36.4 & 39.6 & 0.90 &  0.29(8)  & 0.55 & 0.07(2)  & $-$3.6 \\
NGC 4945   & 21.2 & 37.6 & 39.8 & 0.88 &  0.39(14) & 0.39 & 0.06(2)  & $-$3.8 \\
NGC 5194/5 & 20.2 & 37.4 & 40.0 & 0.84 &  0.33(17) & 0.33 & 0.04(2)  & $-$3.8 \\
NGC 5236   & 20.6 & 36.5 & 39.8 & 0.69 &  0.17(14) & 0.18 & 0.01(1)  & $-$3.7 \\
NGC 5457   & 20.1 & 36.3 & 39.8 & 0.66 &  0.06(8)  & 0.25 & 0.01(1)  & $-$3.4 \\
NGC 6503   & 20.6 & 37.4 & 39.1 & 0.85 &  0.43(3)  & 0.00 & 0.00(0)  & $-$3.6 \\
\tableline
NGC 1395   & 20.3 & 38.4 & 40.2 & 1.00 &  1.00(30) & 0.49 & 0.10(3)  & $-$3.7 \\
NGC 1399   & 20.1 & 37.6 & 40.4 & 0.91 &  0.74(74) & 0.23 & 0.03(3)  & $-$3.6 \\
NGC 3379   & 20.4 & 37.1 & 39.5 & 0.73 &  0.23(8)  & 0.00 & 0.00(0)  & $-$3.8 \\
NGC 4374   & 20.4 & 37.8 & 40.2 & 0.98 &  0.86(24) & 0.63 & 0.04(1)  & $-$3.8 \\
NGC 4472   & 20.2 & 37.7 & 40.3 & 0.96 &  0.87(68) & 0.14 & 0.03(2)  & $-$4.0 \\
NGC 4636   & 20.3 & 37.5 & 39.6 & 0.69 &  0.49(18) & 0.00 & 0.00(0)  & $-$4.1 \\
NGC 4649   & 20.3 & 37.8 & 40.2 & 0.80 &  0.56(55) & 0.07 & 0.01(1)  & $-$3.9 \\
NGC 4697   & 20.3 & 37.6 & 40.1 & 0.86 &  0.58(40) & 0.11 & 0.01(1)  & $-$3.8 \\
NGC 5128   & 20.9 & 36.9 & 40.1 & 0.66 &  0.19(25) & 0.15 & 0.01(1)  & $-$3.8 \\
\enddata

\tablenotetext{~}{Notes on columns: 
(1) Galaxy Name; 
(2) Neutral Hydrogen absorption column of the Milky Way, 
  from the {\sc FTOOLS NH} program (Dickey \& Lockman 1990);
(3) Limiting X-ray luminosity calculated using PIMMS, assuming a detection 
of 10 net counts,
a power-law model with photon index $\Gamma=1.7$, and the Galactic absorption
listed in column 2.  Since most of the galaxies were centered on CCD7 and
most of the point sources were positioned on that chip, we assumed the 
spectral response for CCD7, and the exposure time for CCD7 listed in Table~1,
except for the following galaxies: NGC~3034 (CCD0), NGC~1395 (CCD3),
and NGC~5128 (CCD0).
(4) Observed 
X-ray point-source luminosity, from all point sources
(see Table 1, column 6).  Note that nuclear and near-nuclear 
(within 5$"$ of the NED astrometric position) X-ray point sources have been 
omitted for the following galaxies: 
NGC~3079, NGC~4258, NGC~4579, NGC~4945, NGC~5194, NGC~4374, and NGC~5128.  
Artificial X-ray sources due to saturation along the CCD readout direction 
(``readout stripes'') have been omitted for NGC~4579 and NGC~5128.  
X-ray sources along the ``jet'' in NGC~5128 have also been omitted.
(5,6) Fraction of L$_{XP}$ and N (the number of X-ray point sources listed
in Table 1) from sources with L$_X \ge$ 10$^{38}$ erg~s$^{-1}$.  
The number of sources with L$_X \ge$ 10$^{38}$ erg~s$^{-1}$ is 
listed in parentheses in column 6;
(7,8) Same as columns (5,6) except for 10$^{39}$ erg~s$^{-1}$;
(9) Ratio of L$_{XP}$ to galaxy K$_s$ band luminosity from Table 2.
}

\end{deluxetable}

\clearpage
\begin{deluxetable}{lcccc}
\tabletypesize{\scriptsize}
\tablecaption{Slopes of Cumulative X-ray Luminosity Functions\label{tab4}}
\tablewidth{0pt}
\tablehead{
\colhead{Galaxy} & \colhead{$\bar{L}_X(min)$} & \colhead{$\bar{L}_X(max)$} & \colhead{N} & \colhead{$\gamma$} \\
\colhead{Name} & \colhead{($log$ erg/s)} & \colhead{($log$ erg/s)} & \colhead{} & \colhead{} \\
\colhead{(1)} & \colhead{(2)} & \colhead{(3)} & \colhead{(4)} &
\colhead{(5)} \\

}
\startdata
NGC 1569   & 36.0 & 37.5 &   5 & 0.44 \\
NGC 3034   & 37.5 & 39.5 &  20 & 0.57 \\
NGC 4038/9 & 38.0 & 40.0 &  53 & 0.63 \\
NGC 4449   & 37.0 & 38.4 &  12 & 0.70 \\
NGC 5253   & 36.5 & 37.7 &   8 & 0.92 \\
\tableline
NGC 253    & 37.0 & 38.6 &  36 & 0.75 \\
NGC 628    & 37.5 & 38.6 &  38 & 1.11 \\
NGC 1291   & 37.7 & 38.9 &  31 & 1.16 \\
NGC 2681   & 37.8 & 39.2 &  10 & 0.65 \\
NGC 3079   & 37.5 & 39.2 &  17 & 0.80 \\
NGC 3184   & 37.5 & 38.7 &  20 & 1.15 \\
NGC 3628   & 37.0 & 39.3 &  28 & 0.82 \\
NGC 4244   & 36.6 & 37.5 &   3 & 0.46 \\
NGC 4258   & 37.5 & 38.8 &  23 & 0.79 \\ 
NGC 4314   & 37.7 & 38.2 &   7 & 1.20 \\
NGC 4579   & 38.5 & 40.1 &   5 & 0.47 \\
NGC 4631   & 37.0 & 39.2 &  23 & 0.69 \\
NGC 4736   & 37.2 & 39.1 &  19 & 0.55 \\
NGC 4945   & 37.6 & 39.1 &  22 & 0.70 \\
NGC 5194/5 & 37.7 & 39.1 &  29 & 0.78 \\
NGC 5236   & 37.5 & 39.0 &  34 & 0.91 \\
NGC 5457   & 37.0 & 39.2 &  68 & 0.85 \\
NGC 6503   & 37.5 & 38.7 &   4 & 0.37 \\
\tableline
NGC 1395   & 38.3 & 39.7 &  22 & 1.05 \\
NGC 1399   & 38.0 & 39.4 &  79 & 1.26 \\
NGC 3379   & 37.8 & 39.0 &  12 & 1.07 \\
NGC 4374   & 38.0 & 38.6 &  26 & 1.16 \\
NGC 4472   & 38.5 & 39.2 &  19 & 1.63 \\
NGC 4636   & 38.0 & 38.6 &  22 & 2.36 \\
NGC 4649   & 38.0 & 39.0 &  62 & 1.58 \\
NGC 4697   & 38.0 & 39.1 &  39 & 1.34 \\
NGC 5128   & 38.0 & 39.2 &  28 & 1.28 \\
\enddata

\tablenotetext{~}{Notes on columns: 
(1) Galaxy Name; 
(2,3) The slope $\gamma$ was computed using a weighted least-squares algorithm with
      logarithmic bins in X-ray luminosity.
      $\bar{L}_X(min)$ and $\bar{L}_X(max)$ are the luminosities corresponding
      to (the center of) the minimum and maximum bin used to compute $\gamma$. 
      The bins have width of 0.1 dex.
      $\bar{L}_X(min)$ was chosen to maximize the range in L$_{XP}$ without 
      including very low L$_{XP}$ values, where a completeness problem may
      exist (see Figures 2 and 3).
(4) Total number of X-ray point sources used in the fit;
(5) Slope $\gamma$ of the cumulative XLF
}

\end{deluxetable}

\clearpage

\begin{deluxetable}{lccccll}
\tabletypesize{\scriptsize}
\tablecaption{X-ray Spectral Properties of Luminous HMXBs \label{tab4bis}}
\tablewidth{0pt}
\tablehead{
\colhead{Name} & \colhead{Type} & \colhead{L$_{X}$/10$^{38}$~erg~s$^{-1}$} &
\colhead{Instr.} & \colhead{$\Gamma$} & \colhead{Ref.} &
\colhead{Notes} \\
\colhead{(1)} & \colhead{(2)} & \colhead{(3)} & 
\colhead{(4)} & \colhead{(5)} & \colhead{(6)} & 
\colhead{(7)} \\
}
\startdata
SMC X-2        & Be/T    & 1   & ASCA   &  0.7 & 1 & outburst\\
A~0538$-$66    & Be/T    & 6   & ASCA   & 1.16 & 2 & outburst \\
EXO~2030$+$375 & Be/T    & 1   & EXOSAT & 1.83 & 3 & outburst\\
\hline
SMC X-1        & AcPXP     & 6   & ASCA   & 1.05\tablenotemark{\dag} & 4 & high state \\
LMC X-4        & AcPXP     & 6   & ASCA   & 0.70\tablenotemark{\dag} & 4 & high state \\
LMC X-3        & BHC     & 3   & RXTE   &  1.8\tablenotemark{\dag} & 5 & high state \\
LMC X-1        & BHC     & 2   & RXTE   &  2.9\tablenotemark{\dag} & 5 & high state \\
Cen X-3        & AcPXP     & 1   & BBXRT  &  1.1 & 6 & high state \\
\hline
Cyg X-3        & ?       & 1   &        & $\sim$2 & 7 & high state \\
               &         & 0.4 &        & $\sim$1 & 7 & low state \\
\enddata

\tablenotetext{~}{Notes on columns:
(1,2) Source name, and HMXB type.   
X-ray transients (all in outburst) are listed first (all are Be/Transients), 
then accretion-powered X-ray pulsars (AcPXPs) and accretion-powered
black-hole candidates (BHCs);
(3) X-ray luminosity, as listed in Guseinov et al. (2000);
(4) Instrument used for determining $\Gamma$;
(5) Photon index $\Gamma$ for simple power-law fit;
(6) Reference for $\Gamma$ ([1] Yokogawa et al. 2001, [2] Corbet et al. 1997, 
[3] Parmar et al. 1989, [4] Paul et al. 2002, [5] Nowak et al. 2001, [6] Audley et al. 1996, [7] White \& Holt 1982);
(7) Notes on HMXB state during observation
}

\tablenotetext{\dag}{
Since these sources have complex X-ray spectra, requiring both a 
hard (power-law) and a soft component in when observed with reasonably
S/N, $\Gamma$ was determined by simulating a Chandra ACIS-S X-ray spectrum
with XSPEC, then re-fitting the simulated data with a simple absorbed 
power-law model.  For SMC X-1 and LMC X-4, we used
models III from Tables 1 and 3 from Paul
et al. (2002, respectively) to simulate the spectra, and, for
LMC~X-3 and LMC~X-1, respectively, we used models from Tables
2 and 3 of Nowak et al. (2001).
}

\end{deluxetable}

\clearpage
\begin{deluxetable}{llrllrl}
\tabletypesize{\scriptsize}
\tablecaption{Proportionality Constants for L$_{XP}$ \label{tab6}}
\tablewidth{0pt}
\tablehead{
\colhead{A} & \colhead{B} & \colhead{$\chi_\nu^2$} &
\colhead{$\alpha$} & \colhead{$\beta$} & \colhead{$\chi_\nu^2$} &
\colhead{Sample Description (No. of galaxies)} \\
\colhead{(1)} & \colhead{(2)} & \colhead{(3)} & 
\colhead{(4)} & \colhead{(5)} & \colhead{(6)} & 
\colhead{(7)} \\
}
\startdata
1.26$^{+0.30}_{-0.31}$ & {...}                  & 0.81 & 0.88$^{+0.22}_{-0.21}$ & {...}                  & 0.90 & Elliptical galaxies only (8) \\
1.11$^{+0.11}_{-0.11}$ & 0.31$^{+0.10}_{-0.10}$ & 4.61 & 1.05$^{+0.11}_{-0.11}$ & 0.74$^{+0.15}_{-0.15}$ & 3.79 & All galaxies with FIR flux measurements (30) \\
1.01$^{+0.14}_{-0.14}$ & 0.44$^{+0.14}_{-0.13}$ & 5.53 & 1.09$^{+0.14}_{-0.14}$ & 0.86$^{+0.19}_{-0.19}$ & 4.40 & All galaxies with FIR flux measurements, excepting AGN (24) \\
0.93$^{+0.15}_{-0.15}$ & 0.49$^{+0.14}_{-0.13}$ & 7.21 & 1.25$^{+0.17}_{-0.17}$ & 0.74$^{+0.20}_{-0.18}$ & 5.56 & Only M/I and Spiral galaxies, no AGN (18) \\
\enddata

\tablenotetext{~}{Notes on columns:
(1,2) Dimensionless A and B constants for correlation of L$_{XP}$ with host galaxy luminosities L$_K$ and
  L$_{FIR+UV}$, respectively (see section 5.3, equation 1);
(3) Reduced-$\chi^2$ for the fit
  $=$ $\chi^2$ / (no. of galaxies - no. of parameters);
(4,5) Constants $\alpha$ and $\beta$, for correlation with 
  M and SFR (see equation 2).  Units are
  erg~s$^{-1}$~M$_{\odot}^{-1}$ and 
  erg~s$^{-1}$~(M$_\odot$~yr$^{-1}$)$^{-1}$, respectively;
(6) Reduced-$\chi^2$ for the fit;
(7) Description of the galaxy sample used in the calculation (number of 
galaxies in parentheses)
}

\tablenotetext{~}{
All quoted uncertainties are 90\% confidence for one
parameter free ($\Delta\chi^2 =$ 2.7).  For example, errors in $A$ were
determined by fixing $B$ at its minimum-$\chi^2$ value and scanning over
$A$.
}

\end{deluxetable}

\clearpage

\begin{deluxetable}{lccc}
\tabletypesize{\scriptsize}
\tablecaption{Inferred Properties of Host Galaxies \label{tab5}}
\tablewidth{0pt}
\tablehead{
\colhead{Galaxy} & \colhead{SFR} & \colhead{$log(M/L)_K$} & \colhead{Mass} \\
\colhead{Name} & \colhead{(M$_\odot$~yr$^{-1}$)} & \colhead{($\odot$)} & \colhead{$log$(M$_\odot$)} \\
\colhead{(1)} & \colhead{(2)} & \colhead{(3)} & \colhead{(4)} \\
}
\startdata
NGC 1569   & 0.037 & $-$0.6 &   7.9 \\
NGC 3034   & 9.3   & $-$0.1 &  10.8 \\
NGC 4038/9 & 8.3   & $-$0.4 &  10.7 \\
NGC 4449   & 0.21  & $-$0.4 &   8.9 \\
NGC 5253   & 0.14  & $-$0.5 &   8.4 \\
\tableline
NGC 253    & 2.4   & $-$0.3 &  10.5 \\
NGC 628    & 1.3   & $-$0.4 &  10.0 \\
NGC 1291   & 0.13  & $-$0.2 &  10.7 \\
NGC 2681   & 0.47  & $-$0.3 &  10.3 \\
NGC 3079   & 4.0   & $-$0.3 &  10.5 \\
NGC 3184   & 0.56  & $-$0.4 &   9.8 \\
NGC 3628   & 1.3   & $-$0.3 &  10.3 \\
NGC 4244   & 0.047 & $-$0.7 &   8.4 \\
NGC 4258   & 0.69  & $-$0.3 &  10.4 \\
NGC 4314   & 0.16  & $-$0.2 &  10.1 \\
NGC 4579   & 1.4   & $-$0.2 &  11.1 \\
NGC 4631   & 1.6   & $-$0.5 &   9.8 \\
NGC 4736   & 0.58  & $-$0.2 &  10.4 \\
NGC 4945   & 3.6   & $-$0.3 &  10.6 \\
NGC 5194/5 & 1.8   & $-$0.2 &  10.8 \\
NGC 5236   & 1.2   & $-$0.3 &  10.5 \\
NGC 5457   & 2.5   & $-$0.5 &  10.0 \\
NGC 6503   & 0.20  & $-$0.4 &   9.6 \\
\tableline
NGC 1395   & ...   & $-$0.2 &  11.0 \\
NGC 1399   & ...   & $-$0.1 &  11.2 \\
NGC 3379   & ...   & $-$0.1 &  10.5 \\
NGC 4374   & ...   & $-$0.2 &  11.1 \\ 
NGC 4472   & ...   & $-$0.1 &  11.5 \\
NGC 4636   & ...   & $-$0.1 &  10.9 \\
NGC 4649   & ...   & $-$0.1 &  11.2 \\
NGC 4697   & ...   & $-$0.2 &  11.0 \\
NGC 5128   & 1.7\tablenotemark{a} & $-$0.2 &  10.9 \\
\enddata

\tablenotetext{~}{Notes on columns:
(1) Galaxy name;
(2) Star formation rate, estimated from L$_{FIR+UV}$ (see text section 5.2);
(3) Stellar mass-to-light ratio for the K band, in solar units, 
    calculated from B$-$K (Table 2).  See text section 5.2 for details;
(4) Stellar mass, calculated using L$_K$ from Table 2, and 
    (M/L)$_K$ from column 3.
}

\tablenotetext{a}{%
Since the FIR emission from elliptical galaxies is not necessarily
dominated by
re-radiated
emission from hot stars, it may not be a good predictor of the SFR in
elliptical galaxies.  However, since NGC~5128 does have ongoing star
formation, we have estimated a SFR for it.
}

\end{deluxetable}

\clearpage

\appendix

\tablenum{A1}
\begin{deluxetable}{rrlcrrrrrrrrrrr}
\tabletypesize{\scriptsize}
\tablecaption{Properties of the X-ray Point Sources \label{appendixtable}}
\tablewidth{0pt}
\tablehead{
\colhead{NGC} & \multicolumn{2}{c}{Position} & \colhead{CCD} & \colhead{L$_X$} & 
\colhead{$H$} & \colhead{$\Delta H$} &
\colhead{$M$} & \colhead{$\Delta M$} &
\colhead{$S$} & \colhead{$\Delta S$} &
\colhead{$HM$} & \colhead{$\Delta HM$} &
\colhead{$MS$} & \colhead{$\Delta MS$} \\
\colhead{} & \multicolumn{2}{c}{(J2000)} & \colhead{No.} & \colhead{} & 
\colhead{} & \colhead{} & \colhead{} & \colhead{} \\
\colhead{(1)} & \colhead{(2)} & \colhead{(3)} & \colhead{(4)} & \colhead{(5)} & 
\colhead{(6)} & \colhead{(7)} & \colhead{(8)} & \colhead{(9)} & \colhead{(10)} & 
\colhead{(11)} & \colhead{(12)} & \colhead{(13)} & \colhead{(14)} & \colhead{(15)} \\
}
\startdata
    253  & 00:47:01.4&$-$25:23:23 & 2 & 37.15 & 4.3 & 3.8 & 11.1 & 4.6 & 3.0 & 3.2 & $-$0.4 & 0.40 & $+$0.6 & 0.38 \\ 
         & 00:47:06.7&$-$25:21:26 & 3 & 37.41 & 12.8 & 4.7 & 11.9 & 4.6 & 8.8 & 4.1 & $-$0.0 & 0.27 & $-$0.3 & 0.30 \\ 
         & 00:47:10.2&$-$25:22:33 & 2 & 37.30 & 20.5 & 5.7 & 6.0 & 3.6 & 0.0 & {...} & $+$0.5 & 0.23 & $+$1 & {...} \\ 
         & 00:47:11.1&$-$25:23:31 & 2 & 38.02 & 40.6 & 7.5 & 83.9 & 10.2 & 15.7 & 5.1 & $-$0.4 & 0.10 & $+$0.5 & 0.09 \\ 
         & 00:47:17.6&$-$25:18:11 & 7 & 38.48 & 188.2 & 14.8 & 323.8 & 19.0 & 176.3 & 14.3 & $-$0.3 & 0.05 & $+$0.3 & 0.05 \\ 
         & 00:47:17.6&$-$25:18:27 & 7 & 37.77 & 25.9 & 6.3 & 74.8 & 9.7 & 31.3 & 6.7 & $-$0.5 & 0.11 & $+$0.4 & 0.10 \\ 
         & 00:47:18.5&$-$25:19:14 & 7 & 37.81 & 43.6 & 7.7 & 58.6 & 8.7 & 49.0 & 8.1 & $-$0.1 & 0.11 & $+$0.1 & 0.11 \\ 
         & 00:47:19.6&$-$25:17:21 & 7 & 37.34 & 0.7 & 2.3 & 18.0 & 5.3 & 29.8 & 6.5 & $-$0.9 & 0.24 & $-$0.2 & 0.17 \\ 
         & 00:47:21.0&$-$25:17:47 & 7 & 36.97 & 3.7 & 3.2 & 12.8 & 4.7 & 4.7 & 3.4 & $-$0.6 & 0.33 & $+$0.5 & 0.32 \\ 
         & 00:47:22.6&$-$25:20:51 & 7 & 38.54 & 119.9 & 12.0 & 330.5 & 19.2 & 301.0 & 18.4 & $-$0.5 & 0.05 & $+$0.1 & 0.04 \\ 
         & 00:47:25.2&$-$25:19:45 & 7 & 37.78 & 34.7 & 7.0 & 55.6 & 8.5 & 47.4 & 8.0 & $-$0.2 & 0.12 & $+$0.1 & 0.11 \\ 
         & 00:47:25.4&$-$25:16:43 & 7 & 37.21 & 17.8 & 5.3 & 15.9 & 5.1 & 2.8 & 2.9 & $+$0.1 & 0.22 & $+$0.7 & 0.28 \\ 
         & 00:47:25.5&$-$25:18:04 & 7 & 36.70 & 0.0 & {...} & 3.5 & 3.2 & 8.6 & 4.4 & $-$1 & {...} & $-$0.4 & 0.43 \\ 
         & 00:47:26.5&$-$25:19:14 & 7 & 37.29 & 8.9 & 4.1 & 22.8 & 5.9 & 13.3 & 4.8 & $-$0.4 & 0.21 & $+$0.3 & 0.21 \\ 
         & 00:47:28.0&$-$25:18:20 & 7 & 37.52 & 27.9 & 6.4 & 31.9 & 6.7 & 13.5 & 4.8 & $-$0.1 & 0.15 & $+$0.4 & 0.17 \\ 
         & 00:47:28.4&$-$25:17:09 & 7 & 36.90 & 1.9 & 2.7 & 10.7 & 4.4 & 5.6 & 3.6 & $-$0.7 & 0.38 & $+$0.3 & 0.35 \\ 
         & 00:47:28.6&$-$25:19:23 & 7 & 37.08 & 11.7 & 4.6 & 14.7 & 5.0 & 1.2 & 2.7 & $-$0.1 & 0.26 & $+$0.9 & 0.32 \\ 
         & 00:47:29.9&$-$25:17:38 & 7 & 37.01 & 1.9 & 2.7 & 18.6 & 5.5 & 3.4 & 3.2 & $-$0.8 & 0.24 & $+$0.7 & 0.26 \\ 
         & 00:47:30.9&$-$25:14:50 & 7 & 36.88 & 15.8 & 5.1 & 2.0 & 2.7 & 0.0 & {...} & $+$0.8 & 0.27 & $+$1 & {...} \\ 
\multicolumn{12}{c}{\bf  ... Many Lines Removed (electronic table) ...} \\
         & 14:04:25.1&$+$54:25:51 & 3 & 37.39 & 18.8 & 6.1 & 36.4 & 7.2 & 7.9 & 4.3 & $-$0.4 & 0.17 & $+$0.5 & 0.17 \\ 
         \tableline
    6503  & 17:49:12.5&$+$70:09:31 & 7 & 38.40 & 15.7 & 5.1 & 44.9 & 7.8 & 35.7 & 7.1 & $-$0.5 & 0.14 & $+$0.1 & 0.13 \\ 
         & 17:49:26.4&$+$70:08:40 & 7 & 37.44 & 1.8 & 2.7 & 7.8 & 4.0 & 0.8 & 2.3 & $-$0.6 & 0.47 & $+$0.8 & 0.50 \\ 
         & 17:49:27.9&$+$70:08:37 & 7 & 37.76 & 4.7 & 3.4 & 7.0 & 3.8 & 10.0 & 4.3 & $-$0.2 & 0.43 & $-$0.2 & 0.33 \\ 
         & 17:49:28.8&$+$70:08:33 & 7 & 37.85 & 7.9 & 4.0 & 11.8 & 4.6 & 6.8 & 3.8 & $-$0.2 & 0.30 & $+$0.3 & 0.31 \\ 
         & 17:49:29.1&$+$70:08:44 & 7 & 38.42 & 29.7 & 6.5 & 39.7 & 7.4 & 31.7 & 6.7 & $-$0.1 & 0.14 & $+$0.1 & 0.14 \\ 
         & 17:49:31.7&$+$70:08:20 & 7 & 38.68 & 72.9 & 9.6 & 83.6 & 10.2 & 23.9 & 6.0 & $-$0.1 & 0.09 & $+$0.6 & 0.10 \\ 
         & 17:49:38.1&$+$70:08:32 & 7 & 37.25 & 0.0 & {...} & 0.9 & 2.3 & 5.8 & 3.6 & $-$1 & {...} & $-$0.7 & 0.61 \\ 
\enddata

\tablenotetext{~}{Notes on columns:
(1) Host Galaxy NGC number (see Table 1);
(2,3) X-ray position in J2000 coordinates;
(4) ACIS CCD number.  CCDs 5 and 7 are back-illuminated, and the others are front-illuminated (see section 3.1);
(5) Logarithm of the 0.3$-$8.0 keV X-ray luminosity, assuming galaxy distances listed in Table 2 and exposure times listed in Table 1. See
section 3.1 for details about assumed spectrum;
(6$-$11) Net observed counts and uncertainties for the Hard, Medium and Soft bands ($H$, $M$, and $S$), as described in section 4.3;
(12$-$15) Hardness ratios $HM = (H - S)/(H + S)$ and $MS = (M - S)/(M + S)$ for BI CCD 7, and corresponding uncertainties,
    {\it for front-illuminated CCD number 7}, as described in section 3.1.
Uncertainties are not listed for hardness ratios when no net counts were detected in one of the bands.
}

\end{deluxetable}

\end{document}